\documentstyle[epsfig]{aipproc}

\begin{document}
\title{LOTIS Upper Limits and the Prompt OT from GRB~990123}

\author{
G.~G.~Williams$^*$, D.~H.~Hartmann$^*$, H.~S.~Park$^{\dagger}$,
R.~A.~Porrata$^{\dagger}$, E.~Ables$^{\dagger}$, R.~Bionta$^{\dagger}$,
D.~L.~Band$^{\P}$, S.~D.~Barthelmy$^{\S}$, T.~Cline$^{\S}$,
N.~Gehrels$^{\S}$, D.~H.~Ferguson$^{\ddag}$, G.~Fishman$^{\star}$,
R.~M.~Kippen$^{\star}$, C.~Kouveliotou$^{\star}$, K.~Hurley$^{\sharp}$,
R.~Nemiroff$^{\flat}$, and T.~Sasseen$^{||}$
}

\address{
$^*$Dept. of Physics and Astronomy, Clemson University, Clemson, SC 29634\\
$^{\dagger}$Lawrence Livermore National Laboratory, Livermore, CA 94550\\ 
$^{\P}$Los Alamos National Laboratory, Los Alamos, NM 87545\\
$^{\S}$NASA Goddard Space Flight Center, Greenbelt, MD 20771\\
$^{\ddag}$Dept. of Physics, California State University, Hayward,
CA 94542\\
$^{\star}$NASA Marshall Space Flight Center, Huntsville, AL 35812\\
$^{\sharp}$Space Sciences Laboratory, University of California, Berkeley,
CA 94720\\ 
$^{\flat}$Dept. of Physics, Michigan Technological University, Houghton,
MI 49931\\
$^{||}$Dept. of Physics, University of California, Santa Barbara,
CA 93106
}

\maketitle

\begin{abstract}
GRB~990123 established the existence of prompt optical emission
from gamma-ray bursts (GRBs). The Livermore Optical Transient 
Imaging System (LOTIS) has been conducting a fully automated 
search for this kind of simultaneous low energy emission from
GRBs since October 1996. Although LOTIS has obtained simultaneous,
or near simultaneous, coverage of the error boxes obtained with
BATSE, IPN, XTE, and BeppoSAX for several GRBs, image analysis
resulted in only upper limits. The unique gamma-ray properties of
GRB~990123, such as very large fluence (top 0.4\%) and hard spectrum,
complicate comparisons with more typical bursts. We scale and 
compare gamma-ray properties, and in some cases afterglow properties,
from the best LOTIS events to those of GRB~990123 in an attempt to
determine whether the prompt optical emission of this event is 
representative of all GRBs. Furthermore, using LOTIS upper limits
in conjunction with the relativistic blast wave model, we weakly
constrain the GRB and afterglow parameters such as density of the
circumburster medium and bulk Lorentz factor of the ejecta. 
\end{abstract}

\section*{Introduction}
The ultimate reward for the Gamma-Ray Burst Coordinates Network
(GCN)~\cite{barthelmy98} came when the Robotic
Optical Transient Search Experiment (ROTSE) detected prompt
optical emission from GRB~990123~\cite{akerlof99}.
Although this discovery marks another milestone in 
comprehending the physics of GRBs, bright optical 
transients (OTs) may be the exception rather than 
the rule.  Both LOTIS and ROTSE have unsuccessfully
attempted to detect these predicted flashes on many
occasions~\cite{park97a,park97b,williams98,williams99,schaefer99,kehoe99}.
Although some of the non-detections 
may be attributed to large extinction, GRB~990123 
demonstrated that the progenitor is not always obscured.

\section*{Observations \& Analysis}
During more than 1100 nights of possible observations 
(since October 1996), LOTIS has responded to 127
GCN triggers.  Of these, 68 triggers were 
unique GRB events; a rate of approximately one unique GRB 
event every 16.5 days.  The quality of the LOTIS 
``coverage'' for a given event depends on five factors:
observing conditions, LOTIS response time, difference
between the initial and final coordinates, size of the
final error box, and the duration of the GRB.
Table~\ref{A-09-table1} lists 13 events for which 
LOTIS achieved good coverage. 

\begin{table}[b]
\caption{LOTIS GRB events with good coverage and predictions for
the scaled magnitudes of the prompt optical emission.} 
\label{A-09-table1}
\begin{tabular}{ccdddddd}
 Date & BATSE & $F_{p}$ (64 ms) & $F_{p}$ (1024 ms) & $S/10^{-7}$ & $m_{F;64}$ & $m_{F;1024}$ & $m_{S}$ \\
         & Trig. & \multicolumn{2}{c}{($\gamma$ cm$^{-2}$ s$^{-1}$)} & (erg cm$^{-2}$)   &                    &                      &                    \\
\tableline
990123 & 7343 & 16.96 & 16.41 & 3000. & 9.0  & 9.0  & 9.0   \\
\tableline
961017 & 5634 & 4.22  & 1.98  & 5.07  & 10.5 & 11.3 & 15.9  \\
961220 & 5719 & 1.93  & 1.60  & 18.11 & 11.4 & 11.5 & 14.5  \\
970223 & 6100 & 19.41 & 16.84 & 968.  & 8.9  & 9.0  & 10.2  \\
970714 & 6307 & 1.89  & 1.32  & 17.09 & 11.4 & 11.7 & 14.6  \\
970919 & 6388 & 1.10  & 0.77  & 22.49 & 12.0 & 12.3 & 14.3  \\
971006 & 6414 & 2.08  & 1.79  & 258.  & 11.3 & 11.4 & 11.7  \\
971227 & 6546 & 3.32  & 2.11  & 9.25  & 10.8 & 11.2 & 15.3  \\
990129 & 7360 & 5.88  & 4.99  & 585.  & 10.2 & 10.3 & 10.8  \\
990308 & 7457 & 2.02  & 1.26  & 164.  & 11.3 & 11.8 & 12.2  \\
990316 & 7475 & 3.87  & 3.67  & 529.  & 10.6 & 10.6 & 10.9  \\
990413 & 7518 & 3.78  & 2.57  & 68.13 & 10.6 & 11.0 & 13.1  \\
990803 & 7695 & 16.99 & 12.19 & 1230. & 9.0  & 9.3  & 10.0  \\
990918 & 7770 & 5.69  & 3.17  & 25.21 & 10.2 & 10.8 & 14.2  \\
\end{tabular}
\end{table}

First we compare GRB~990123 with the LOTIS upper limits 
to test whether the flux of the prompt optical
emission scales with some gamma-ray property.
Here and throughout the analysis we neglect extinction
effects.  The first row in Table~\ref{A-09-table1} 
lists the properties of GRB~990123~\cite{briggs99,akerlof99}.
The columns display the UTC date of the burst, 
the BATSE trigger number, the 64~ms and 1024~ms peak
fluxes~(50~-~300~keV), and the gamma-ray fluence~($>$20~kev) 
of each event.  The last three columns are the scaled 
magnitudes,
\begin{equation}
        m_{\rm GRB} = m_{\rm GRB990123} - 2.5%
	\log \left( \frac{X_{\rm GRB}}{X_{\rm GRB990123}} \right),
\label{A-09-eq1}
\end{equation}
where $m_{\rm GRB990123} = 9.0$, the peak magnitude 
of GRB~990123, and $X_{\rm GRB}$ and $X_{\rm GRB990123}$
are the peak flux or fluence values for those events.

The LOTIS sensitivity varies depending on observing
conditions but in general a conservative limiting 
magnitude is $m \approx 11.5$ prior to March~1998
(upgrade to cooled CCD) and $m \approx 14.0$ 
following that date.  Table~\ref{A-09-table1} shows
that the scaled prompt optical emission for both peak
flux and fluence is often brighter than the LOTIS upper
limits which suggests that these simple relationships
are not valid.

Briggs {\it et al.}~\cite{briggs99} show that 
the optical flux measured during GRB~990123 
is not consistent with an extrapolation of the 
burst spectrum to low energies.  However 
Liang {\it et al.}~\cite{liang99} point out
that the extrapolated tails rise and fall
with the optical flux.  A low energy enhancement 
would produce an upward break which might account
for the measured optical flux during GRB~990123.
It is important to determine if there is a low
energy upturn in the spectrum since it would 
establish whether or not the optical and gamma-ray 
photons are produced by the same electron distribution.
The LOTIS upper limits can be used to constrain
a low energy enhancement assuming it is common to 
all GRBs.

\begin{figure}[b!] 
\centerline{\epsfig{file=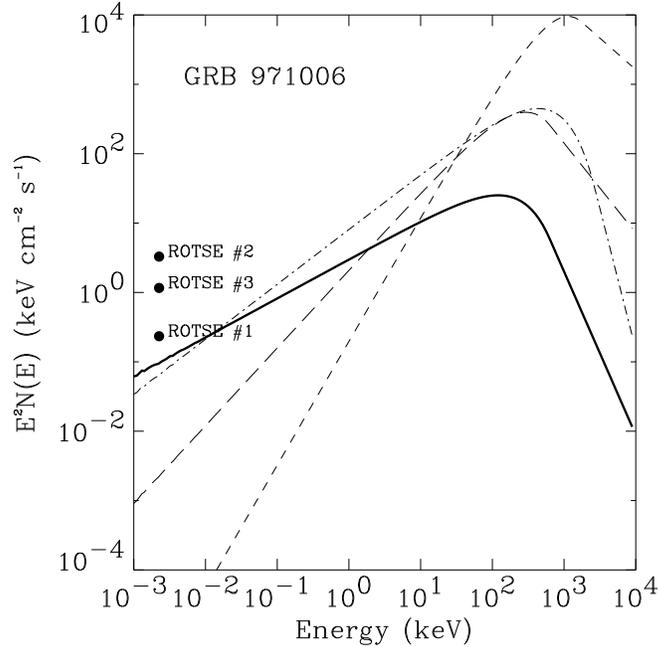,height=3.5in,width=3.5in}}
\vspace{10pt}
\caption{Extrapolated spectrum of GRB~971006 during the LOTIS observation
and GRB~990123 during ROTSE detections.}
\label{A-09-fig2}
\end{figure}

For the events listed in Table~\ref{A-09-table1}
we fit the gamma-ray spectra during the LOTIS
observations to the Band functional form~\cite{band93}. 
In a few cases the low energy extrapolation is
near the LOTIS upper limit.  The solid
line in Figure~\ref{A-09-fig2} shows the Band 
fit to GRB~971006 and its extrapolation to low
energies.  Fits to the spectra of GRB~990123
during the first (short dash), second (dash-dot), and 
third (long dash) ROTSE observations and 
the corresponding ROTSE detections (filled circles)
are also shown.  The extrapolation of GRB~971006,
predicts an $m \approx 12.4$ optical flash.
Even a slight upward break in the spectrum would
have produced a detectable OT.  We conclude 
that the LOTIS upper limits support the
hypothesis that the low energy emission is
produced by a different electron distribution
than the high energy emission.

Finally we attempt to use the LOTIS upper limits
and the external reverse shock model 
to constrain the physical properties of the
GRB blast wave.  Sari and Piran~\cite{sari99}
show that the fraction of the energy which 
gets emitted in the optical band depends on 
the values of the cooling frequency and the 
characteristic synchrotron frequency.  For the
external reverse shock these frequencies are
given by
\begin{equation}
        {{\nu}_c} = 8.8 \times 10^{15} {\rm Hz} {\left( \frac{{\epsilon}_B}%
        {0.1} \right)}^{-3/2} {E_{52}}^{-1/2} {n_1}^{-1} {t_A}^{-1/2},
\label{A-09-eq2}
\end{equation}
\begin{equation}
        {{\nu}_m} = 1.2 \times 10^{14} {\rm Hz} {\left( \frac{{\epsilon}_e}%
        {0.1} \right)}^{2} {\left( \frac{{\epsilon}_B} {0.1} \right)}^{1/2}%
        {\left( \frac{{\gamma}_0} {300} \right)}^{2} {n_1}^{1/2},
\label{A-09-eq3}
\end{equation}
where ${\epsilon}_e$ and ${\epsilon}_B$ are the
fraction of equipartition energy in the electrons
and magnetic field, $E_{52}$ is the total energy
in units of $10^{52}$ erg, $n_1$ is the density of
circumburster medium in cm$^{-3}$, ${\gamma}_0$ is the 
initial Lorentz factor, and $t_A$ is the duration
of the emission in seconds.

\begin{figure}[t!] 
\centerline{\epsfig{file=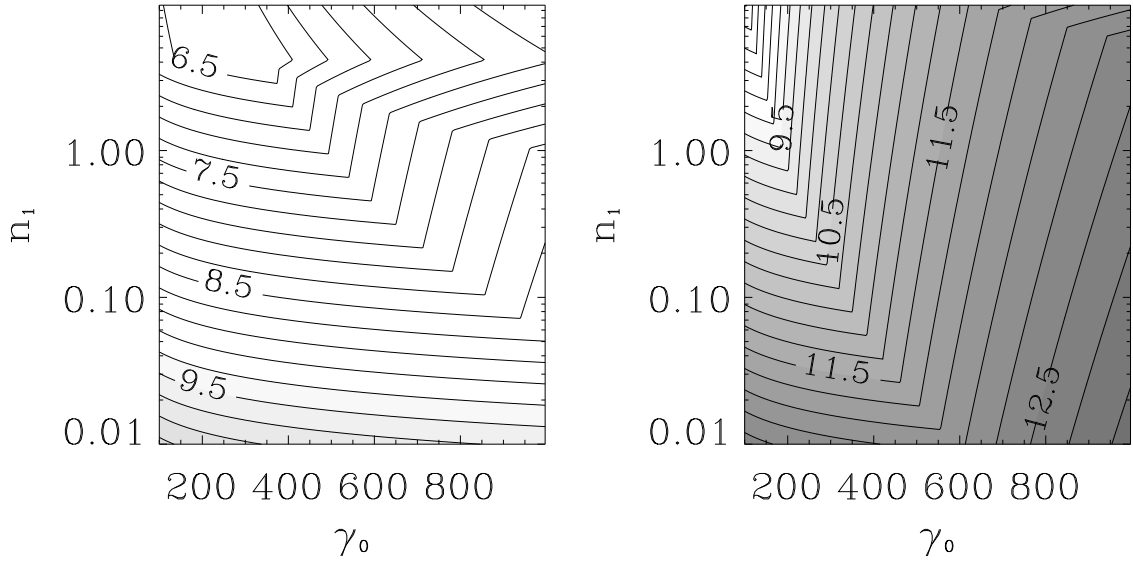,height=3.0in}}
\vspace{10pt}
\caption{Predicted magnitude of the prompt optical flash for $E_{52} = 3.5$,
${\epsilon}_e = 0.12$, and ${\epsilon}_B = 0.089$ (left panel) and 
$E_{52} = 0.53$, ${\epsilon}_e = 0.57$, and ${\epsilon}_B = 0.0082$ (right
panel).}
\label{A-09-fig1}
\end{figure}

Sari and Piran assume the frequency
dependencies modify the fluence of a moderately
strong GRB, i.e. $10^{-5}$~erg~cm$^{-2}$.  In this
analysis we compare the afterglow properties of GRB~970508
found by Wijers and Galama~\cite{wijers99} to those 
found by Granot {\it et al.}~\cite{granot99}.
Therefore we use a fluence of 
$3.1 \times 10^{-6}$~erg~cm$^{-2}$ emitted over
the entire LOTIS integration time of $t_A = 10.0$ s.
The index of the electron power-law distribution
is set to ${\rm p} = 2.2$.

Figure~\ref{A-09-fig1} shows contour plots of
the predicted magnitude of the prompt OT 
for GRB~970508 as a function of $n_1$ and ${\gamma}_0$.
GRB~970508 could not be observed 
by LOTIS or ROTSE since it occurred during the
day.  Values of $E_{52} = 3.5$, ${\epsilon}_e = 0.12$,
and ${\epsilon}_B = 0.089$ from Wijers and Galama are
used in the left panel and values of $E_{52} = 0.53$,
${\epsilon}_e = 0.57$, and ${\epsilon}_B = 0.0082$
from Granot {\it et al.} are used in the right panel.
The right panel demonstrates the effect of altering
the total energy and the distribution of energy to 
the electrons and the magnetic field.  The smaller 
values of $E_{52}$ and ${\epsilon}_B$ shift the 
contours to the upper left while the larger 
${\epsilon}_e$ steepens the breaks in the contours.
The increased shading corresponds to a decreasing
detection probability.  However for nearly all 
values of $n_1$ and ${\gamma}_0$ shown the predicted
OT could have been detected by the upgraded
LOTIS system.

Wijers and Galama find a circumburster medium 
density of $n_1 = 0.030$ which predicts an
$m = 9.0 - 9.5$ optical flash nearly
independent of the initial Lorentz factor. 
Granot {\it et al.} find a considerably higher vlaue
of $n_1 = 5.3$, which predicts an $m = 8.7 - 12.4$
OT which is very dependent on the initial Lorentz
factor.  The LOTIS upper limits mildly favor 
the GRB blast wave values determined by 
Granot {\it et al.} since dim OTs are predicted 
over a larger range of initial Lorentz factors.

\end{document}